\documentclass[12pt]{article}
\usepackage[utf8]{inputenc}
\usepackage{titlesec}
\usepackage{epsfig}
\usepackage{graphicx}
\usepackage{graphics}
\usepackage{amssymb}
\usepackage{amsmath}
\usepackage{amsfonts}
\usepackage{color}
\usepackage{textcomp}
\usepackage{latexsym}
\usepackage{cite}
\usepackage{titletoc}
\usepackage{tikz}
\usepackage{xcolor}
\usetikzlibrary{3d}
\usepackage{csquotes}
\usepackage{caption}
\usetikzlibrary{calc} 
\usetikzlibrary{angles,quotes}
\usepackage{csquotes}
\titleformat{\section}{\normalfont\Large\bfseries}{\thesection.}{1em}{}
\titlecontents{section}[0em]{\bfseries}{\thecontentslabel. }{}{\titlerule*[1pc]{.}\contentspage}
\usepackage[margin=2.5cm,footskip=0.6cm]{geometry}
\begin{document}

\begin{center}
{\Large{\bf Velocity Resetting of Inertial Run-and-Tumble Particles in Non-Newtonian Media: Velocity Distribution, Diffusion and First-Passage Time} } \\
\ \\
by \\
Subhanker Howlader, Sayantan Mondal and Prasenjit Das\footnote{prasenjit.das@iisermohali.ac.in} \\
Department of Physical Sciences, Indian Institute of Science Education and Research -- Mohali, Knowledge City, Sector 81, SAS Nagar 140306, Punjab, India. \\
\end{center}

\begin{abstract}
\noindent We study the dynamics of an athermal inertial run-and-tumble particle moving through a non-Newtonian medium in $d=1$, where the particle’s velocity $v$ is reset to zero at a constant rate $r$. The drag force from the non-Newtonian medium is represented by a nonlinear velocity-dependent function $g(v)$. The run-and-tumble dynamics is modeled by a symmetric dichotomous noise with strength $\Sigma$ and flipping rate $\lambda$. We begin with the Fokker–Planck~(FP) equation for the velocity distribution $P(v,t)$ of the particle. In the presence of resetting, however, the FP equation does not yield a closed-form solution even in the steady state. We therefore compute the steady-state velocity distribution $P_s(v)$ directly from particle trajectories and compare it with the numerical solution of the FP equation, finding good agreement between the two approaches. For sufficiently large $r$, $P_s(v)$ shows cusp-like singularity at $v=0$ and the particles display diffusive motion at long times. The effective diffusion coefficient $D_{\rm eff}$ decays as $r^{-2}$ in the large-$r$ regime. These results hold irrespective of the specific form of $g(v)$, and the values of $\lambda$ and $\Sigma$. However, the mean first-passage time exhibits a strong dependence on the nature of the medium as the resetting rate $r$ is varied. In shear-thickening media, there exists an optimal resetting rate that minimizes the time required to reach the target velocity $v_t$. In contrast, no such optimal resetting rate is observed in shear-thinning media.
\end{abstract}

\newpage
\section{Introduction}
Stochastic resetting is a dynamical process in which one or more dynamical variables are intermittently interrupted and reset to a predefined state according to a specified protocol~\cite{p2c1mresnm11,p2c1mresnm011,PKE16,EMS20,ES2025}. In recent years, it has emerged as a powerful mechanism for controlling nonequilibrium processes~\cite{p2c1besgaprr20,p2c1faisant21jsm}. Problems of this type were initially introduced in the context of search processes and have since been found useful in areas ranging from computer science~\cite{p2c1viswanathan99}, for the development of algorithms and information retrieval, to the chemical sciences, where they are applied to reaction dynamics~\cite{p2c1harris12,BRH2022,CBKPRH2025}. In the biological sciences, the concept of resetting has proved valuable for understanding animal foraging strategies~\cite{p2c1bell90,p2c1turchin98}, protein target search on DNA molecules~\cite{p2c1facdiffdna09}, as well as human behavior and population dynamics~\cite{p2c1bartumeus09,p2c1physforaging11}. More broadly, stochastic resetting has attracted interest in applied mathematics, physics, and computer science due to its conceptual simplicity and analytical tractability~\cite{p2c1lubyzuc93,p2c1gMontazec02,p2c1stojkoski22rsta,p2c1rotbart15pre,p2c1belan18prl,PCBProcA2025}. Since the resetting mechanism generates a probability current toward the resetting point or region, the system typically evolves toward a nonequilibrium steady state~\cite{SPAP2025}.

The paradigmatic example of stochastic resetting is the one-dimensional Brownian motion subjected to Poissonian resetting to a fixed spatial point, as studied by Evans and Majumder~\cite{p2c1mresnm11}. Starting from the microscopic dynamics, they derived the Fokker-Planck~(FP) equation for the probability density function $P_r(x,t|x_0,t_0)$. For a finite resetting rate $r$, the solution of the FP equation can be obtained using Laplace transforms, leading to a nonequilibrium stationary state $P_r^*(x|x_0,t_0)$ for the particle position. Furthermore, they showed that resetting introduces a characteristic length scale $l=\sqrt{D/r}$, where $D$ is the diffusion constant, and that $P_r^*(x|x_0,t_0)$ depends explicitly on $l$. They also calculated the mean first-passage time to a fixed point in space, which becomes finite for $r\ne0$ and attains a minimum at an optimal resetting rate $r^*$. This problem has been extended to cases involving a time-dependent resetting rate $r(t)$, non-Poissonian resetting characterized by power-law waiting-time distributions between resetting events, and arbitrary stochastic processes such as L\'evy flights and fractional Brownian motion~\cite{p2c1stojkoski22rsta,p2c1humphries12,p2c1ariel15,APSR17,p2c1shokaku20,p2c1guinard21}. Moreover, the study of resetting has been extended to multi-particle systems~\cite{BKPPRE19,AAA25,BMP25PRL} as well as to extended systems~\cite{JKP22}, such as fluctuating interfaces described by the KPZ equation~\cite{SSG14,p2c1silvafragoso18}.

There exist only a few experimental studies on stochastic resetting~\cite{Roichman24,GLPHMG25,KSOHLLC25}. The first such experimental study was carried out by Tal-Friedman \textit{et al.}, who investigated the dynamics of a colloidal particle under stochastic resetting using holographic optical tweezers~\cite{p2c1talfriedman20}. Their experimental results for the steady-state position distribution, first-passage properties, and the cost of resetting in the steady state show good agreement with analytical predictions. Later, Vatash and Roichman studied the dynamics of many interacting colloidal particles under stochastic resetting in order to elucidate the competing effects of particle interactions on the resulting steady-state distribution~\cite{RY25}. Ginot and Bechinger studied the resetting dynamics of a particle in a viscoelastic fluid~\cite{FC26}. The memory effects induced by the viscoelastic medium give rise to elastic restoring forces that oppose the resetting protocol and hinder efficient exploration. Recently, Antonov \textit{et al.} studied the dynamics of an inertial active particle subject to solid friction, where the particle experiences a non-Newtonian drag force in the absence of resetting~\cite{Antonov24}.

All the studies discussed above consider overdamped dynamics of particles with position resetting only. However, velocity resetting protocols are equally physically relevant when inertia plays a significant role~\cite{DGupta19,PSingh20,Santra2025}. For example, in the Andersen thermostat, the temperature of a many-particle system is controlled by resetting particle velocities to a Gaussian distribution at a predefined rate~\cite{H80, MD1987}. A perfectly inelastic collision of a particle with a substrate is equivalent to resetting the particle’s velocity to zero~\cite{NT04}. Intermittent stopping during the motion of flying foragers, such as bees visiting flowers, also corresponds to velocity resetting, as the velocity is reset to zero. Similarly, stop-and-go locomotion observed in macroscopic living organisms represents a form of velocity resetting. In such cases, constituents intermittently halt their motion completely, for instance to conserve energy or to scan for predators~\cite{DLKRLM01,WTADDLK04,FB2009,MSDEWH10,ADWJGJG10,TEHPKLDM11}. Examples include lizards climbing trees and fish resting in water. In all these scenarios, the positions of the particles remain unaffected. The first theoretical study of velocity resetting for an inertial particle moving in a Newtonian fluid is due to Olsen and L\"owen~\cite{OL2024}. They described this protocol as \textit{partial stochastic resetting}, in which the velocity is reset while the position remains unchanged. Using Fourier–Laplace transforms, they obtained general expressions for the effective diffusion and drift coefficients as functions of the resetting rate. At late times, the system exhibits normal diffusion, although velocity resetting generically suppresses velocity correlations. Extensions of this study to stochastic resetting with refractory periods show that such protocols can induce anomalous diffusion~\cite{KH24}. Finally, all their analytical results are validated by numerical simulations. 

In this paper, we study the velocity resetting dynamics of an inertial run-and-tumble particle (IRTP) moving in a non-Newtonian medium. Unlike the Newtonian media, where the viscous drag force $g(v)$ is linearly proportional to the particle velocity $v$, i.e., $g(v)\sim v$, non-Newtonian media exert a drag force that depends nonlinearly on the velocity. For example, $g(v)\sim \tan(v)$ mimics shear-thickening medium, since $g(v)$ increases rapidly with increasing $v$. Such non-Newtonian behavior is common in a wide variety of complex fluids and biological environments found in nature. Examples include colloidal suspensions, polymer solutions, paints, cellular fluids, mucus, and blood, all of which often exhibit shear-thinning or shear-thickening behavior rather than a simple linear viscous response \cite{afnhfb08,esjbcssk14,nsrflgg19,jcpacwwkcfp23}. In biological systems, both cells and bacteria frequently move through non-Newtonian media and may experience effective velocity resetting due to intrinsic stop-and-go dynamics or inelastic collisions with substrates and surrounding structures~\cite{KGS23}. For nonlinear $g(v)$, the Fourier–Laplace transform approach to solving the FP equation becomes inapplicable, making the calculation of time-dependent or steady-state velocity distribution functions and their moments in the presence of resetting difficult. Therefore, we address the problem using two complementary approaches: (i) numerical solutions of the FP equation, and (ii) direct simulations of the particle’s equations of motion, which we then compare.

Given this background, the paper is organized as follows. Section~\ref{sec2} provides the details of the modeling, the FP equation, and the theoretical framework for calculating the mean first-passage time. The simulation setup and numerical results are discussed in Sec.~\ref{sec3}. Finally, we summarize our findings in Sec.~\ref{sec4}.

\section{Details of Modeling}\label{sec2}
Consider an IRTP of mass $m$ moving in a non-Newtonian medium in $d=1$. The viscous drag force acting on the particle depends on its speed $v$ and is modeled by a nonlinear function $g(v/v_0)$, where $v_0$ is the characteristic velocity. The intrinsic active force responsible for the run-and-tumble motion of the particle is modeled by a symmetric dichotomous noise $\zeta(t)$~\cite{WR83,J84,I06}. So, the Langevin equation that describe the particle motion is given by
\begin{eqnarray}
    \label{p2c4eqn1}
    m\frac{dv}{dt} = -A g\left(\frac{v}{v_0}\right) + \zeta.
\end{eqnarray}
Here, $A$ represents the amplitude of the drag force. Active force $\zeta(t)$ flips between two discrete values, $\pm\nu$,  with probability $\Gamma dt$ during a time interval $dt$, where $\Gamma$ is the constant transition rate. So, $\zeta(t)$ has zero mean and obeys the following autocorrelation function:
\begin{eqnarray}  
\label{p2c4eqn2}  
\langle \zeta(t) \zeta(t^{\prime}) \rangle = \nu^2 \exp\left(-2 \Gamma |t - t^{\prime}|\right).  
\end{eqnarray}

Next, we reset the velocity $v(t)$ to zero at random times drawn from an exponential distribution with a uniform rate $r\geq0$. Accordingly, over an infinitesimal time interval $dt$, the evolution of $v(t)$ is given by
\begin{equation}
\label{p2c4eqn3}
v(t+dt)=
\begin{cases}
\text{updated according to Eq.~(\ref{p2c4eqn1}) with probability } 1-rdt,\\
0 \quad \text{with probability } rdt.
\end{cases}
\end{equation}
Here, we assume that the flipping of the active forces and the resetting events are statistically independent~\cite{OL2024}.

We nondimensionalize Eq.~(\ref{p2c4eqn1}) by rescaling velocity and time as
\begin{eqnarray}  
\label{p2c4eqn4}  
\quad v = v_0 v^\prime \quad \text{and} \quad t = \frac{m v_0}{A} t^\prime. 
\end{eqnarray}
The primed variables $v'$ and $t'$ are the dimensionless velocity and time, respectively. Substituting $v$ and $t$ into Eq.~(\ref{p2c4eqn1}) and dropping the primes, we obtain the nondimensional form of the Langevin equation:
\begin{eqnarray} 
\label{p2c4eqn5}
\frac{dv}{dt} = -g(v) + \eta(t). 
\end{eqnarray}
The rescaled active force $\eta(t)$ exhibits the following properties:
\begin{eqnarray}  
\label{p2c4eqn6}  
\langle \eta(t) \rangle = 0  \quad \text{and} \quad \langle \eta(t) \eta(t^{\prime}) \rangle = \Sigma^2 \exp\left(-2\lambda|t - t^{\prime}|\right),  
\end{eqnarray}
Here, $\Sigma = \nu / A$ and $\lambda = m v_0 \Gamma / A$, respectively, represent the scaled amplitude and scaled flipping rate of $\eta(t)$. Finally, the scaled resetting rate is defined as
\begin{eqnarray}  
\label{p2c4eqn7}  
r^{\prime} = r\frac{mv_0}{A},  
\end{eqnarray}
and, we subsequently drop the prime.

\begin{table}[h]
\caption{Choice of $g(v)$}
\centering
\begin{tabular}{|c|c|}
\hline
Shear-Thickening & Shear-Thinning \\ 
\hline
$g(v)=\tan(v)$  & $g(v)=\tanh(v)$  \\  
\hline
\end{tabular}
\label{tab1}
\end{table}
Table~\ref{tab1} lists the choices $g(v)$ that characterize the shear-thickening and shear-thinning responses of the medium~\cite{A15,TA23,ICTS19,MD25,showsmpdas25}. Next, we determine the domain of admissible velocities for a fixed amplitude $\Sigma$ of $\eta(t)$ by analyzing the fixed points of the velocity dynamics governed by Eq.~(\ref{p2c4eqn5}). For the shear-thickening medium, setting $\eta(t) = +\Sigma$ yields the upper bound $+v_u$ of $v$:
\begin{eqnarray}  
\label{p2c4eqn8}  
-\tan(v_u)+\Sigma=0 \Rightarrow v_u = \tan^{-1}\Sigma.  
\end{eqnarray}
Similarly, $\eta(t) = -\Sigma$ gives the lower bound $v_l=-\tan^{-1}\Sigma$. Consequently, for the shear-thickening medium, the velocity is restricted to 
\begin{eqnarray}  
\label{p2c4eqn9}
v\in (-\tan^{-1}\Sigma,+\tan^{-1}\Sigma).
\end{eqnarray}
One can determine that, for the shear-thinning medium, the velocity is instead bounded within
\begin{eqnarray}  
\label{p2c4eqn10}
v\in (-\tanh^{-1}\Sigma,+\tanh^{-1}\Sigma).
\end{eqnarray}
Therefore, $\Sigma$ must be less than unity for the shear-thinning medium, while it can take arbitrary values for the shear-thickening medium.

\subsection{Fokker-Planck Equation}\label{sec2.1}
We write the Fokker-Planck~(FP) equations corresponding to the Langevin dynamics~(\ref{p2c4eqn5}) in the presence of resetting~(\ref{p2c4eqn3})~\cite{MRESNM18,PCB20,C85,H96,V08}. We define $P_{\pm}^r(v, t)$ as the conditional probability density for the particle to have velocity $v$ at time $t$, given the active force is $\pm\Sigma$ and the constant rate of resetting is $r$. The FP equations for these probability densities are given by
\begin{eqnarray}
\label{p2c4eqn11}
\frac{\partial}{\partial t} P_{+}^r(v, t) = \frac{\partial}{\partial v} \left[g(v) - \Sigma\right]P_{+}^r(v, t) - rP_{+}^r(v, t) + \frac{r}{2}\delta(v) + \lambda\left[P_{-}^r(v, t) - P_{+}^r(v, t)\right],\\
\label{p2c4eqn12}
\frac{\partial}{\partial t} P_{-}^r(v, t) = \frac{\partial}{\partial v} \left[g(v) + \Sigma\right]P_{-}^r(v, t) - rP_{-}^r(v, t) + \frac{r}{2}\delta(v) + \lambda\left[P_{+}^r(v, t) - P_{-}^r(v, t)\right].
\end{eqnarray}
In Eqs.~(\ref{p2c4eqn11}) and (\ref{p2c4eqn12}), the first term on the right-hand side represent the deterministic evolution of $P_{\pm}^r(v, t)$ in $v$ space under the influence of the drag force $g(v)$, and active force $+\Sigma$ or $-\Sigma$. The term $-rP_{\pm}^r(v,t)$ corresponds to the loss of probability due to resetting for $v\ne0$, while $\frac{r}{2}\delta(v)$ represents the gain of probability at $v=0$ resulting from the resetting of the velocity to $v=0$. The last term accounts for the redistribution of probability densities due to flipping of the active force between the two states at a fixed velocity $v$.

Next, we define the total probability density $P^r(v,t)$ and an auxiliary function $Q^r(v,t)$ as
\begin{eqnarray}  
\label{p2c4eqn13}
&&P^r(v,t) = P_+^r(v,t) + P_-^r(v,t),\\
\label{p2c4eqn13a}
&&Q^r(v,t) = \lambda[P_+^r(v,t) - P_-^r(v,t)].
\end{eqnarray}
So, $P^r(v,t)$ represents the probability density of finding the particle with velocity $v$ at time $t$ and resetting rate $r$, irrespective of the state of the active force. Since resetting drives the system into a nonequilibrium steady state, we will primarily focus on the steady-state distribution $P^r(v,t)$, which we denote by $P_s^r(v)$.

Adding Eqs.~(\ref{p2c4eqn11}) and (\ref{p2c4eqn12}), we obtain the evolution equation for $P^r(v, t)$ as
\begin{equation}
\label{p2c4eqn11a}
\frac{\partial}{\partial t}P^r(v, t) = \frac{\partial}{\partial v}\left[g(v)P^r(v, t)\right] -\frac{\Sigma}{\lambda}\frac{\partial}{\partial v}Q^r(v, t) - rP^r(v, t) + r\delta(v).
\end{equation}
Again, subtracting Eqs.~(\ref{p2c4eqn11}) and (\ref{p2c4eqn12}) yields the evolution equation for $Q^r(v, t)$ as
\begin{equation}
\label{p2c4eqn12a}
\frac{\partial}{\partial t}Q^r(v, t)=\frac{\partial}{\partial v}\left[g(v)Q^r(v, t)\right] - \lambda\Sigma\frac{\partial}{\partial v}P^r(v, t) - rQ^r(v, t) - 2\lambda Q^r(v, t).
\end{equation}
Our earlier studies with a nonlinear form of $g(v)$ showed that a time-dependent solution for $P^r(v, t)$ is not possible even at $r=0$~\cite{showsmpdas25,MD25}. Therefore, we focus on obtaining the steady-state solution $P_s^r(v)$, as done previously. In the steady state, Eqs.~(\ref{p2c4eqn11a}) and (\ref{p2c4eqn12a}) reduces to
\begin{eqnarray}
\label{p2c4eqn11b}
\frac{d}{dv}\left[g(v)P_s^r(v)\right] -\frac{\Sigma}{\lambda}\frac{d}{dv}Q_s^r(v) - rP_s^r(v) + r\delta(v)=0,\\
\label{p2c4eqn12b}
\frac{d}{dv}\left[g(v)Q_s^r(v)\right] - \lambda\Sigma\frac{d}{dv}P_s^r(v) - (2\lambda+r)Q_s^r(v)=0.
\end{eqnarray}
The solution of Eq.~(\ref{p2c4eqn12b}) is given by
\begin{eqnarray}
\label{p2c4eqn12c}
Q_s^r(v)=\frac{\exp\left({(2\lambda +r)\int\frac{dv}{g(v)}}\right)}{g(v)}\lambda\Sigma\int\frac{1}{g(v)}\frac{dP_s^r(v)}{dv}\exp\left({-(2\lambda +r)\int\frac{dv}{g(v)}}\right)dv.
\end{eqnarray}
Here, we set the integration constant to zero by imposing the boundary condition $Q_s(v)=0$ for $v>v_u$ and $v<v_l$. This is because particle's velocity is restricted to $v\in(v_l,v_u)$~\cite{showsmpdas25}. Finally, substituting $Q_s^r(v)$ into Eq.~(\ref{p2c4eqn11b}) yields an integro-differential equation for $P_s^r(v)$, which is not analytically solvable for nonlinear $g(v)$ and $r\ne0$. Therefore, we focus on obtaining $P_s^r(v)$ by numerically solving the coupled FP equations~(\ref{p2c4eqn11}) and (\ref{p2c4eqn12}). In our earlier studies, we analytically obtained $P_s^r(v)$ up to a normalization constant for nonlinear $g(v)$ functions for $r=0$ given in Table~\ref{tab1}.

\subsection{Theory For Mean First-Passage Time}\label{sec2.2}
Here we present a theory for the mean first-passage time (MFPT) $\tau_r(v_0)$. It is defined as the time taken by a particle, starting with velocity $v_0$ at $t=0$, to reach a target velocity $v_t$ in the presence of resetting. We introduce $S_r^\pm(v_0,t)$ as the survival probability up to time $t$, i.e., the probability that the particle’s velocity has not reached $v_t$ by time $t$, given that it started from $v_0$ with $\eta(0)=\pm\Sigma$. Similarly, $S_r^{\pm}(v_0, t + dt)$ represents the probability that the particle survives up to $t+dt$. Our aim is to derive a dynamical equation for $S_r^\pm(v_0,t)$~\cite{MRESNM18,MGSNMGS25}.

Over an infinitesimal time interval $dt$, resetting occurs with probability $rdt$, while no resetting occurs with probability $1-r dt$. Similarly, tumbling occurs with probability $\lambda dt$, and no tumbling occurs with probability $1-\lambda dt$. There are four possibilities those govern the evolution of $S_r^\pm(v_0,t)$: (i) No resetting and no tumbling: probability $(1 - rdt)(1 - \lambda dt)$; (ii) Tumbling but no resetting: probability $(1 - rdt)\lambda dt = \lambda dt - \mathcal{O}(dt^2)$; (iii) Resetting but no tumbling: probability $rdt(1 - \lambda dt) = rdt - \mathcal{O}(dt^2)$; (iv) Both resetting and tumbling: probability $rdt\lambda dt = r\lambda dt^2$. In all these cases, we assume that only the velocity is reset, and that the tumbling of the active force occurs independently of the resetting. If there is no tumbling and resetting occurs from time $t=0$ to time $t=dt$ and the particle starts with $\eta(0)=+\Sigma$, then the particle's velocity becomes $v_0 + [g(v) + \Sigma]dt$ in time $dt$. If the particle tumbles to $\eta(dt) = -\Sigma$ without resetting, then its new velocity becomes $v_0 + [g(v)-\Sigma]dt$ in time $dt$. Therefore, for the $+\Sigma$ state,
\begin{align}
\label{neqn1}
\!\!S_r^+(v_0,t+dt)&=(1-rdt)(1-\lambda dt)S_r^+\left(v_0+\left[g(v_0)+\Sigma\right]dt,t\right) \nonumber\\ &+ (1-rdt)\lambda dtS_r^-\left(v_0+\left[g(v_0)-\Sigma\right]dt,t\right)+(1-\lambda dt)rdtS_r^+(v_r,t)+\mathcal{O}(dt^2).
\end{align}
The first two terms on the RHS of Eq.~\eqref{neqn1} represent cases (i) and (ii) respectively, while the third term corresponds to the case (iii). In all cases, after a duration $dt$, the particle still needs to survive for a time $t$. Expanding the terms on the RHS and retaining terms up to order $dt$, we obtain
\begin{align}
\label{neqn2}
S_r^+(v_0,t+dt)&=S_r^+(v_0,t)+\left[g(v_0)+\Sigma\right]dt\,\partial_{v_0}S_r^+(v_0,t)-(\lambda+r)dt\,S_r^+(v_0,t)+\lambda dt\,S_r^-(v_0,t) \nonumber\\ & + rdt\,S_r^+(v_r,t).
\end{align}
Taking the $dt\to 0$ limit, we obtain the dynamical equation for $S_r^+(v_0,t)$ as
\begin{align}
\label{neqn3}
\partial_t S_r^+(v_0,t)=\left[g(v_0)+\Sigma\right]\partial_{v_0}S_r^+(v_0,t)-(\lambda+r)S_r^+(v_0,t) +\lambda S_r^-(v_0,t) +rS_r^+(v_r,t).
\end{align}
Similarly, the dynamical equation for $S_r^-(v_0,t)$ can be written as
\begin{align}
\label{neqn4}
\partial_t S_r^-(v_0,t)=\left[g(v_0)-\Sigma\right]\partial_{v_0}S_r^-(v_0,t)-(\lambda+r)S_r^-(v_0,t) +\lambda S_r^+(v_0,t) +rS_r^-(v_r,t).
\end{align}

Next, the conditional mean first-passage times $\tau_r^\pm(v_0)$ are given by
\begin{equation}
\label{neqn5}
\tau_r^\pm(v_0) = \int_0^\infty S_r^\pm(v_0,t)\,dt .
\end{equation}
Integrating Eqs.~\eqref{neqn3} and \eqref{neqn4} over time, and using the boundary conditions $S_r^\pm(v_0,0)=1$ and $S_r^\pm(v_0,\infty)=0$, we obtain
\begin{equation}
\label{neqn6}
\int_0^\infty \partial_t S_r^\pm(v_0,t)\,dt = -1.
\end{equation}
Therefore, Eqs.~\eqref{neqn3} and \eqref{neqn4} after integration over time from 0 to $\infty$ yield
\begin{eqnarray}
\label{neqn26}
&& \left[g(v_0)+\Sigma\right]\partial_{v_0}\tau_r^+(v_0)-(\lambda+r)\tau_r^+(v_0)+\lambda\tau_r^-(v_0)+r\tau_r^+(v_r)=-1,\\
\label{neqn27}
&& \left[g(v_0)-\Sigma\right]\partial_{v_0}\tau_r^-(v_0)+\lambda\tau_r^+(v_0)-(\lambda+r)\tau_r^-(v_0)+r\tau_r^-(v_r)=-1.
\end{eqnarray}
These are the MFPT equations for the state-preserving resetting.

The total mean first-passage time is defined as
\begin{equation}
\label{neqn9}
\tau_r(v_0) = \frac{\tau_r^+(v_0)+\tau_r^-(v_0)}{2}.
\end{equation}
Using Eqs.~\eqref{neqn26} and \eqref{neqn27}, we obtain the following equation for $\tau_r(v_0)$:
\begin{equation}
\label{neqn10}
\left[\Sigma^2-g^2(v_0)\right]\tau_r''(v_0)+2g(v_0)\left[\lambda+r-g'(v_0)\right]\tau_r'(v_0)=\left[g'(v_0)-(2\lambda+r)\right]\left[1-r\left\{\tau_r(v_0)-\tau_r(v_r)\right\}\right].
\end{equation}
To proceed further, one needs a complete solution of Eq.~\eqref{neqn10}. It can, in principle, be solved self-consistently to obtain the MFPT with resetting, $\tau_r(v_0)$. However, this requires two boundary conditions that are not readily available, making the analytical solution cumbersome. In contrast, the boundary conditions for $\tau_r^\pm(v_0)$ are known and are given by (assuming $v_t>0$): (i) $\tau_r^+(v_0)=0$ at $v_0=v_t$ as the particle will reach the target velocity with the active force $+\Sigma$; (ii) $\lim_{v_0\to v_l} \left[g(v_0)-\Sigma\right]\partial_{v_0}\tau_r^-(v_0)=0$, where $v_l=-\tan^{-1}\Sigma$ or $v_l=-\tanh^{-1}\Sigma$ depending on the form of $g(v)$. Therefore, we obtain $\tau_r(v_0)$ numerically by solving Eqs.~\eqref{neqn26} and \eqref{neqn27}, and compare the results with the particle-based simulations in Sec. \ref{sec3}. 
Further, we emphasize that a quadrature expression for $\tau_r(v_0)$ is possible for a general $g(v)$, while closed-form solutions are available only for specific choices of $g(v)$ in the absence of resetting\cite{ICTS19,MRESNM18,MGSNMGS25}.

\section{Simulation Details and Results}\label{sec3}
We numerically solve Eq.~(\ref{p2c4eqn5}) using an Euler discretization scheme, incorporating the resetting rule in Eq.~(\ref{p2c4eqn3}) to update the particle velocity. The discretized version of Eq.~(\ref{p2c4eqn5}) is
\begin{eqnarray} 
\label{p2c4eqn14}
v(t+dt) = v(t) + dt[-g(v) + \eta(t)]. 
\end{eqnarray}
We also update the particle position $x(t)$ using $\dot{x}(t)=v(t)$; its discretized form is
\begin{eqnarray} 
\label{p2c4eqn15}
x(t+dt) = x(t) + v(t)dt. 
\end{eqnarray}
Here, $dt$ denotes the discrete time step and we set $dt=0.001$. We generate the dichotomous noise $\eta(t)$ following the algorithm outlined in Ref.~\cite{CE06}. The probability $q$ that $\eta(t)$, starting from state $+\Sigma$(or $-\Sigma$), flips to state $-\Sigma$(or $+\Sigma$) in time interval $dt$ is given by
\begin{eqnarray}
\label{p2c4eqn16}
q=\frac{1}{2}\{1-\exp(-2\lambda dt)\}.
\end{eqnarray}
Next, we generate a random number $w\in [0,1]$ using a uniform random number generator. When $q>w$, $\eta(t)$ flips. From these simulations, we compute the mean-squared velocity $\langle v^2(t)\rangle$, $P_s(v)$ and mean-squared displacement $\langle x^2(t)\rangle$. At $t=0$, we begin by placing all the particles at the origin with zero velocity. All the distribution functions and statistical results are obtained by using $10^6$ particles. Here, we emphasize that resetting $v(t)$ effectively acts as a partial resetting of $x(t)$ and hence $x(t)$ evolves continuously.

In addition, we numerically solve the coupled FP equations~(\ref{p2c4eqn11}) and (\ref{p2c4eqn12}) using a \textit{finite-difference upwind scheme} to integrate them forward in time~\cite{p2c1pmazumdarcomp21}. At $t=0$, particles are assumed to be located at the origin $v=0$ of the $v$-space. The probability density functions $P_{+}^r(v,0)$ and $P_{-}^r(v,0)$ are then initialized as delta distributions located at $v=0$. In all our simulations, we have fixed the values $P_{+}^r(0,0) = 0.5/\Delta v$ and $P_{-}^r(0,0) = 0.5/\Delta v$, respectively, so that the total probability is normalized. Here, $\Delta v$ denotes the discretized mesh size in the $v$ space. Any positive values of $P_{+}^r(0,0)$ and $P_{-}^r(0,0)$ may be chosen, provided that their sum yields a normalized total probability. For the shear thickening medium, we discretize $v$-space into $2000$ grid points for $\Sigma=3$, with the grid spanning from $-\tan^{-1}\Sigma$ to $+\tan^{-1}\Sigma$. For the shear thinning case, the 
$v$-space is discretized into $1000$ grid points for $\Sigma=0.6$, with the grid extending from $-\tanh^{-1}\Sigma$ to $+\tanh^{-1}\Sigma$. To ensure probability conservation, we employ the no-flux boundary conditions. We use $dt=10^{-4}$ for solving the FP equations, which provides sufficient numerical stability.

\begin{figure}[ht!]
\centering
\includegraphics*[width=0.70\textwidth]{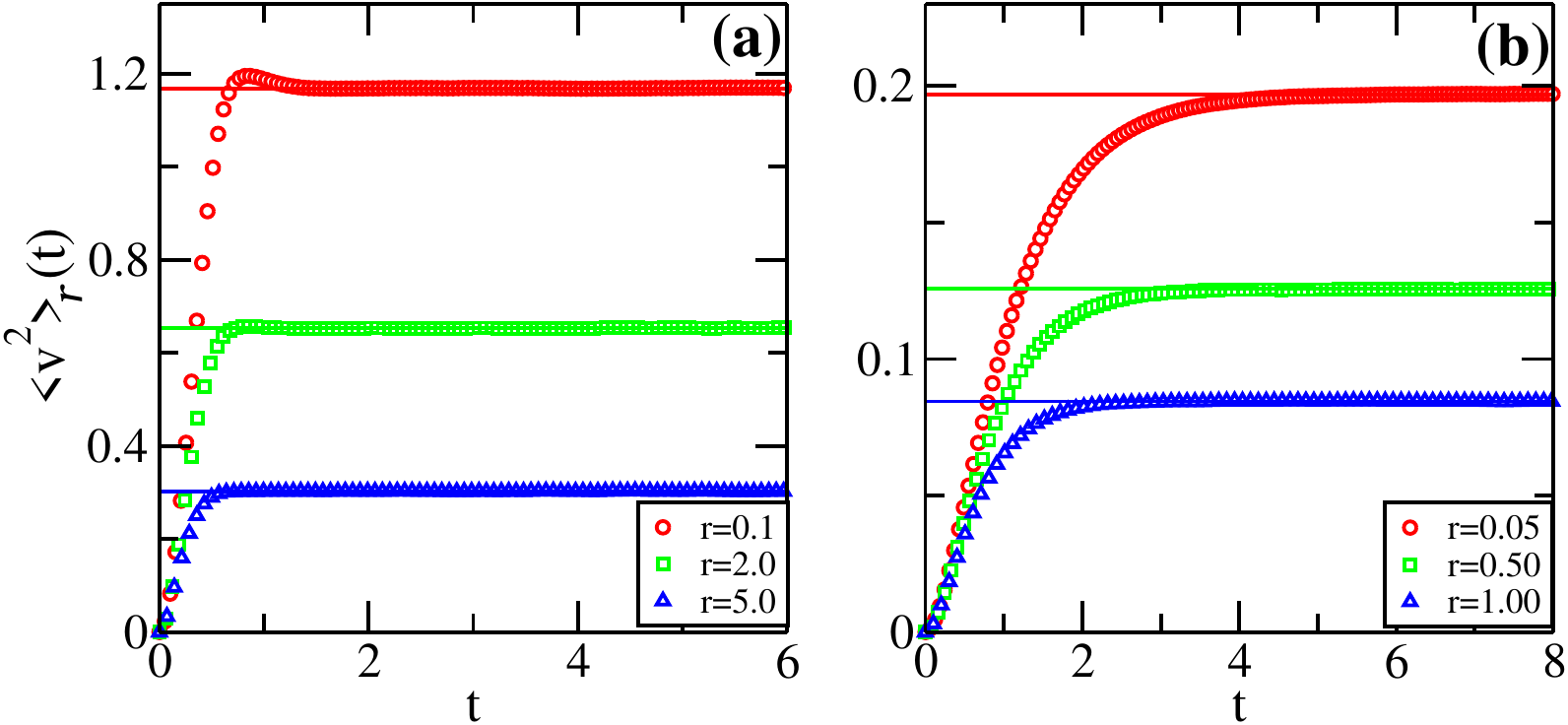}
\caption{\label{p2c4fig1} Plot of mean-squared velocity $\langle v^2\rangle_r(t)$, as a function of time $t$ for $\lambda=0.5$ at different resetting rates, as indicated. The data points represent numerical results obtained by solving Eq.~(\ref{p2c4eqn5}), while the solid lines denote the steady-state mean-squared velocity, $\langle v^2\rangle_r^{\rm ss}$, computed from Eq.~(\ref{p2c4eqn17}). Data points and solid lines of the same color correspond to the same parameter values, with panel (a) showing the shear-thickening case for $\Sigma=3$ and panel (b) the shear-thinning case for $\Sigma=0.6$.}
\end{figure}

Figure~\ref{p2c4fig1} shows the time evolution of $\langle v^2\rangle_r(t)$ for different resetting rates $r$ in shear-thickening [Fig.~\ref{p2c4fig1}(a)] and shear-thinning [Fig.~\ref{p2c4fig1}(b)] media, as mentioned. After an early transient, $\langle v^2\rangle_r(t)$ approaches a steady-state $\langle v^2\rangle_r^{\rm ss}$ value that depends on $r$ in both cases, with all other parameters held fixed. As $r$ increases, $\langle v^2\rangle_r^{\rm ss}$ decreases because more frequent resetting repeatedly drives the particles back to zero velocity. We also compute $\langle v^2\rangle_r^{\rm ss}$ using
\begin{eqnarray}
\label{p2c4eqn17}
\langle v^2\rangle_r^{\rm ss}=\frac{\int_{v_l}^{v_u}dv v^2P_s^r(v)}{\int_{v_l}^{v_u}dv P_s^r(v)},
\end{eqnarray}
where $P_s^r(v)$ is taken from the solutions of FP equations. The numerical results for $\langle v^2\rangle_r^{\rm ss}$ are in excellent agreement with those obtained from Eq.~(\ref{p2c4eqn17}) at long times. Moreover, these results provide a rough estimate of the timescale over which the system relaxes to the steady state.

\begin{figure}[ht!]
\centering
\includegraphics*[width=0.70\textwidth]{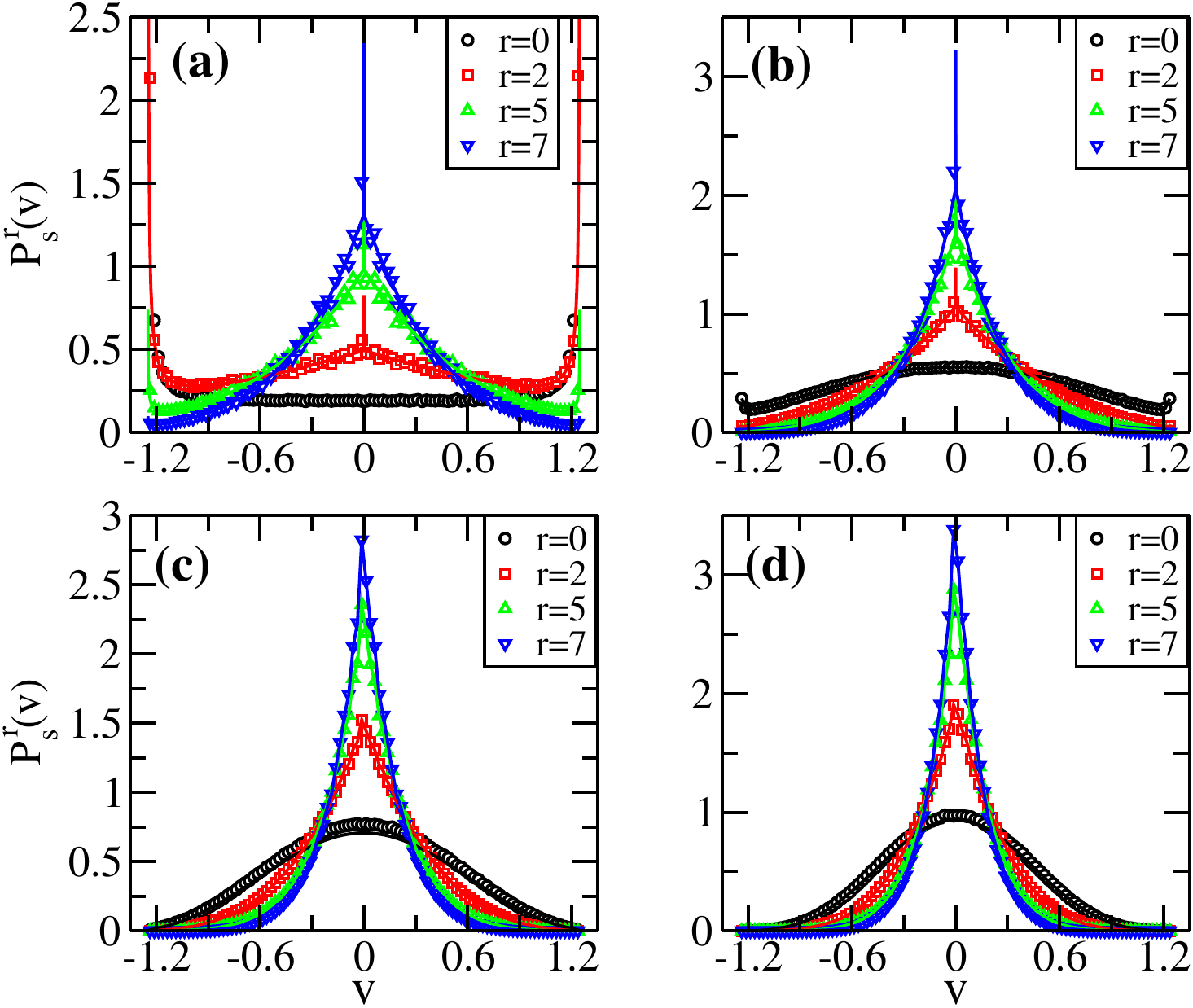}
\caption{\label{p2c4fig2} Plot of steady state velocity distribution functions $P_s^r(v)$ for the shear-thickening medium with $\Sigma=3$ and different values of resetting rates $r$, as indicated. The flipping rates of the active force are (a) $\lambda=0.90$, (b) $\lambda=7$, (c) $\lambda=15$, and (d) $\lambda=25$. The data points correspond to the numerical results obtained by solving particle's equation of motion, while the solid lines represent $P_s(v)$ obtained by numerically solving FP equations at later times.}
\end{figure}

Figure~\ref{p2c4fig2} shows the plots of $P_s^r(v)$ for the shear thickening medium for different $\lambda$ and $r$ with $\Sigma=3$. In the absence of resetting ($r=0$), we denote $P_s^r(v)$ by $P_s(v)$. An earlier study by Howlader \textit{et al.} found the following form of $P_s(v)$~\cite{showsmpdas25}: 
\begin{eqnarray}
\label{Neqn17}
P_s(v) = \frac{N_1}{\Sigma^2 -\tan^2(v)}  
\left[ \frac{\Sigma^2- \tan^2(v)}{1 + \tan^2(v)} \right]^{\frac{\lambda}{(1+\Sigma^2)}},
\end{eqnarray}
where $N_1$ is the normalization constant. It shows several intriguing transitions as $\lambda$ is varied for a fixed $\Sigma$: (i) for $\lambda \leq1$, $P_s(v)$ has a minimum at $v=0$ while diverges to infinity at $v=\pm\tan^{-1}\Sigma$; (ii) for $1<\lambda < 1 + \Sigma^2$, $P_s(v)$ attains a maximum at $v = 0$ and develops a pair of minima at $v=\pm\tan^{-1}\left(\sqrt{\lambda-1}\right)$, while still diverging at $v=\pm\tan^{-1}\Sigma$; (iii) for $(1 + \Sigma^2)<\lambda<2(1+\Sigma^2)$, $v = 0$ becomes the global maximum of $P_s(v)$ and $P_s(v)\rightarrow 0$ at $v=\pm\tan^{-1}\Sigma$ with cusp-like behavior; and (iv) for $\lambda>2(1+\Sigma^2)$, both $P_s(v)$ and it's derivative $P_s^\prime(v)$ vanishes at $v=\pm\tan^{-1}\Sigma$, while $v = 0$ continues to be the maximum of $P_s(v)$. The black solid lines and points in all subfigures of Fig.~\ref{p2c4fig2} represent the results in the absence of resetting.

In Fig.~\ref{p2c4fig2}(a), we plot $P_s^r(v)$s for $\lambda=0.9$ and different values of $r$, as mentioned. The chosen value $\lambda = 0.9$ corresponds to case (i). As $r$ is gradually increased, the minimum in $P_s^r(v)$ at $v = 0$ disappears, and a cusp-like singularity emerges instead due to resetting of velocities at $v = 0$. Simultaneously, a pair of minima appear near the boundaries of the velocity domain located at $v=\pm\tan^{-1}\Sigma$, while the value of $P_s^r(v)$ decreases at the boundaries in a manner similar to case (ii). Upon further increasing $r$, cusp-like singularity at $v=0$ becomes sharper, the pair of minima near $v=\pm\tan^{-1}\Sigma$ vanishes, and $P_s^r(v)$ decays to zero at $v=\pm\tan^{-1}\Sigma$ with a cusp-like singularity, and eventually approaches zero with a horizontal slope, analogous to cases (iii) and (iv), respectively, in the absence of resetting ($r=0$). These rich transitions in $P_s^r(v)$ arise from the complex interplay between resetting and the active force: the former tends to localize the particle at $v=0$, while the latter drives it toward the boundaries of the velocity domain. Figure~\ref{p2c4fig2}(b) shows the corresponding plot for $\lambda=7$, which falls within case (ii) with $r=0$. Introducing a nonzero resetting rate $r$ transforms the smooth central peak into a cusp-like singularity and eliminates the pair of minima present in the absence of resetting. As in Fig.~\ref{p2c4fig2}(a), the tails of $P_s^r(v)$ eventually approach zero for sufficiently large $r$, with the decay occurring through an intermediate regime characterized by cusp-like behavior.
In Figs.~\ref{p2c4fig2}(c) and \ref{p2c4fig2}(d), we plot $P_s^r(v)$ for $\lambda=15$ and $\lambda=25$, which correspond to cases (iii) and (iv) with $r=0$, respectively. In both situations, increasing $r$ transforms the smooth single maximum at $v=0$ into a cusp-like singularity. In all cases, the cusp becomes progressively sharper as $r$ increases, and the tails of $P_s^r(v)$ ultimately decay to zero for sufficiently large $r$. Additionally, the numerical results from direct particle simulations are in excellent agreement with those obtained from solving the corresponding FP equation for the same set of parameters.

\begin{figure}[ht!]
\centering
\includegraphics*[width=0.70\textwidth]{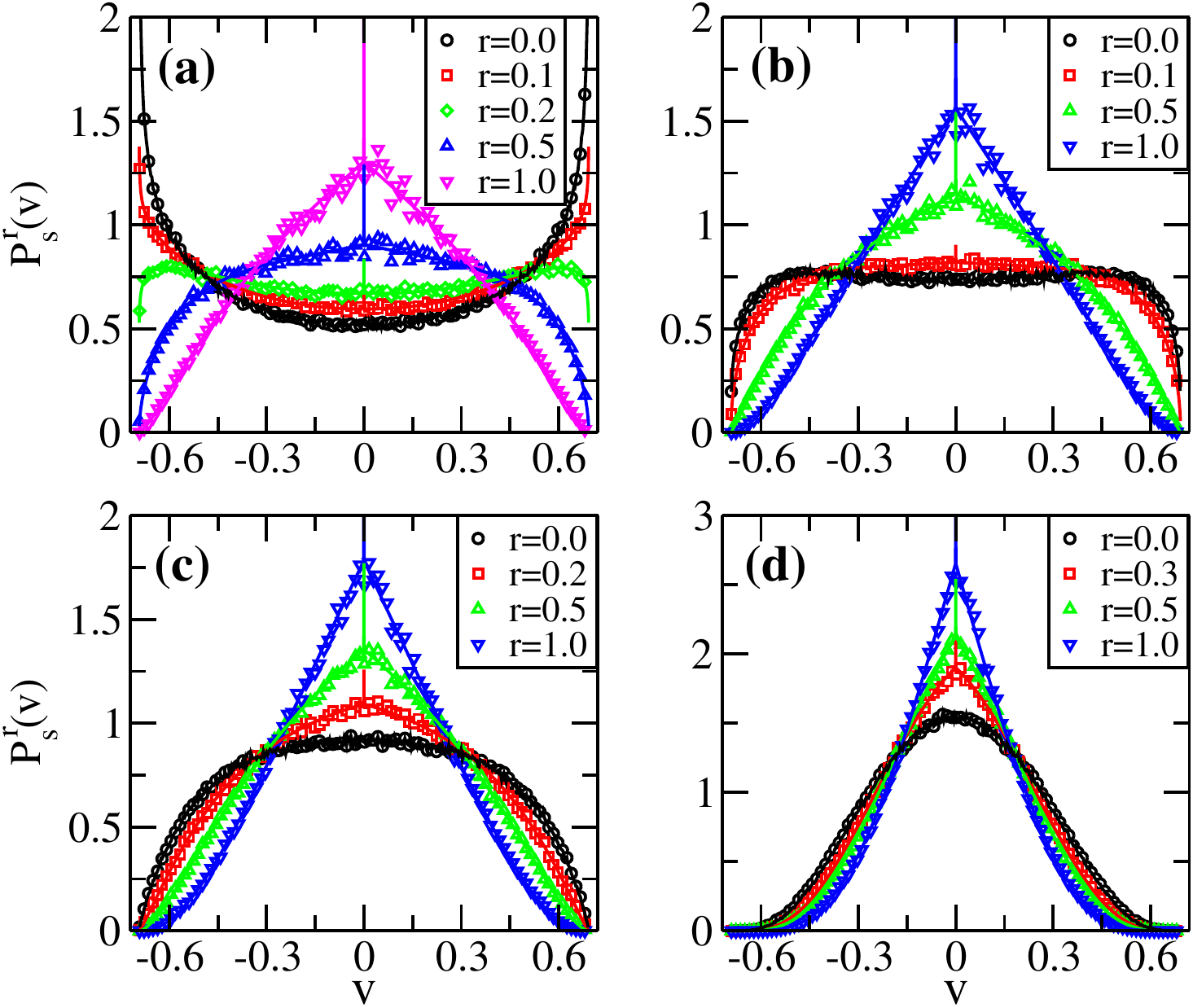}
\caption{\label{p2c4fig3} Figure analogous to Fig.~\ref{p2c4fig2} for the shear-thinning medium with $\Sigma=0.6$ and different values of resetting rates $r$, as indicated.. The flipping rates of the active force are (a) $\lambda=0.5$, (b) $\lambda=0.85$, (c) $\lambda=1.2$, and (d) $\lambda=3$. Other details are the same as those given in the caption of Fig.~\ref{p2c4fig2}.}
\end{figure}

Next, we consider the shear-thinning case for different values of $\lambda$ and $r$ at a fixed $\Sigma$, as shown in Fig.~\ref{p2c4fig3}. Similar to the shear thickening case, $P_s^r(v)$ also exhibits multiple transitions when the model parameters are varied. For $r=0$, Das and Mondal obtained the following form of $P_s(v)$ for the shear-thinning case~\cite{MD25}:
\begin{eqnarray}
\label{apeqn18a}
P_s(v) = \frac{N_2}{\Sigma^2 -\tanh^2(v)} 
\left| \frac{\Sigma^2- \tanh^2( v)}{1 - \tanh^2(v)} \right|^{\frac{\lambda}{(1-\Sigma^2)}}, 
\end{eqnarray}
where $N_2$ is the normalization constant. It shows several transitions as $\lambda$ is varied for a fixed $\Sigma$: (a) when $\lambda\leq 1-\Sigma^2$, $P_s(v)$ has a minimum at $v=0$, and it diverges at the edges of the velocity domain located at $v=\pm \tanh^{-1}{\Sigma}$; (b) for $1-\Sigma^2 < \lambda< 1$, $P_s(v)$ develops a global minima at $v=0$ and a pair of maxima at  $v=\pm \tanh^{-1}\left(1-\lambda\right)$; (c) in the range $(1-\Sigma^2)<\lambda<2(1-\Sigma^2)$, a global maximum emerges at $v=0$. Additionally, $P_s(v)$ approaches zero, and its derivative $P_s^{\prime}(v)$ diverges at $v=\pm \tanh^{-1}{\Sigma}$, indicating cusp singularity at $v=\pm \tanh^{-1}{\Sigma}$; (d) for $\lambda > 2(1-\Sigma^2)$ both $P_s(v)$ and $P^{\prime}(v)$ approaches to zero at $v=\pm \tanh^{-1}{\Sigma}$, with a single maximum at $v=0$. The black solid lines and points in all subfigures of Fig.~\ref{p2c4fig3} correspond to the results without resetting.

Figure~\ref{p2c4fig3}(a) shows plots of $P_s^r(v)$ at different values of $r$ for $\lambda=0.5$. The chosen value of $\lambda$ corresponds to case (a). As we increase $r$, several transitions take place. Around $r=2$, two local maxima emerge near $v=\pm \tanh^{-1}{\Sigma}$, replacing the earlier divergence at $v=\pm \tanh^{-1}{\Sigma}$. Simultaneously, a cusp-like singularity develops at $v=0$. Upon further increasing $r$, the cusp-like singularity at $v=0$ becomes sharper, while $P_s^r(v)$ approaches zero with a diverging slope, similar to case (c) for intermediate $r$, and with a horizontal slope, as mentioned in case (d), for larger $r$ at $v=\pm \tanh^{-1}{\Sigma}$. Figure~\ref{p2c4fig3}(b) shows the plot of $P_s^r(v)$ for $\lambda=0.85$ at different values of $r$, where $r=0$ corresponds to the reset-free result (case (b)). As $r$ increases, the two maxima merge into a single maximum at $v=0$, which subsequently develops into a cusp-like singularity. With further increase in $r$, the cusp-like singularity at $v=0$ becomes more pronounced. Simultaneously, the behavior of $P_s^r(v)$ at $v=\pm \tanh^{-1}{\Sigma}$ changes from a divergent slope, $P_s'(\pm \tanh^{-1}{\Sigma})\rightarrow \infty$, to a horizontal slope, $P_s'(\pm \tanh^{-1}{\Sigma})\rightarrow 0$. In Figs.~\ref{p2c4fig3}(c) and \ref{p2c4fig3}(d), we plot $P_s^r(v)$ for $\lambda=1.2$ and $\lambda=3$, which correspond to cases (c) and (d), respectively, with $r=0$. In Fig.~\ref{p2c4fig3}(c), $P_s^r(v)$ initially displays a cusp-like singularity near $v=\pm\tanh^{-1}{\Sigma}$, but as $r$ is further increased, this cusp-like singularity vanishes and the tail of the distribution function approaches zero horizontally at $v=\pm\tanh^{-1}{\Sigma}$. In contrast, in Fig.~\ref{p2c4fig3}(d), $P_s^r(v)$ vanishes with a horizontal slope for all values of $r$. In both cases, as $r$ is gradually increased, the cusp at $v=0$ becomes increasingly prominent. Again, the numerical data obtained from the particle’s equation of motion are fully consistent with the solution of the FP equation.

From the results shown in Figs.~\ref{p2c4fig2} and \ref{p2c4fig3}, we conclude that, for a fixed $\lambda$, $P_s^r(v)$ undergoes multiple resetting-induced transitions in both shear-thinning and shear-thickening media. In the large-$\lambda$ regime, the velocity of the IRTP fluctuates predominantly around $v=0$, resulting in a Gaussian-like distribution when $r=0$. The introduction of a nonzero resetting rate transforms these distributions into exponentially decaying function: $P_s^r(v)\sim \exp[-\alpha(r)|v|]$, with a cusp at the resetting point ($v=0$) in both cases. Here, $\alpha(r)$ is a positive function of $r$ whose specific form depends on $g(v)$. In contrast, for small $\lambda$, the two media exhibit qualitatively distinct behaviors, both in the presence and absence of resetting [cf. Figs.~\ref{p2c4fig2}(a) and \ref{p2c4fig3}(a)]. Although our analysis is carried out in velocity space, similar transitions have also been reported in position space for reset-free~($r=0$) overdamped systems. In particular, earlier studies of run-and-tumble particles in confining potentials have demonstrated analogous transitions in the position distribution function. These transitions arise from the competition between active driving forces and confinement-induced relaxation~\cite{ICTS19}.

\begin{figure}
\centering
\includegraphics*[width=0.80\textwidth]{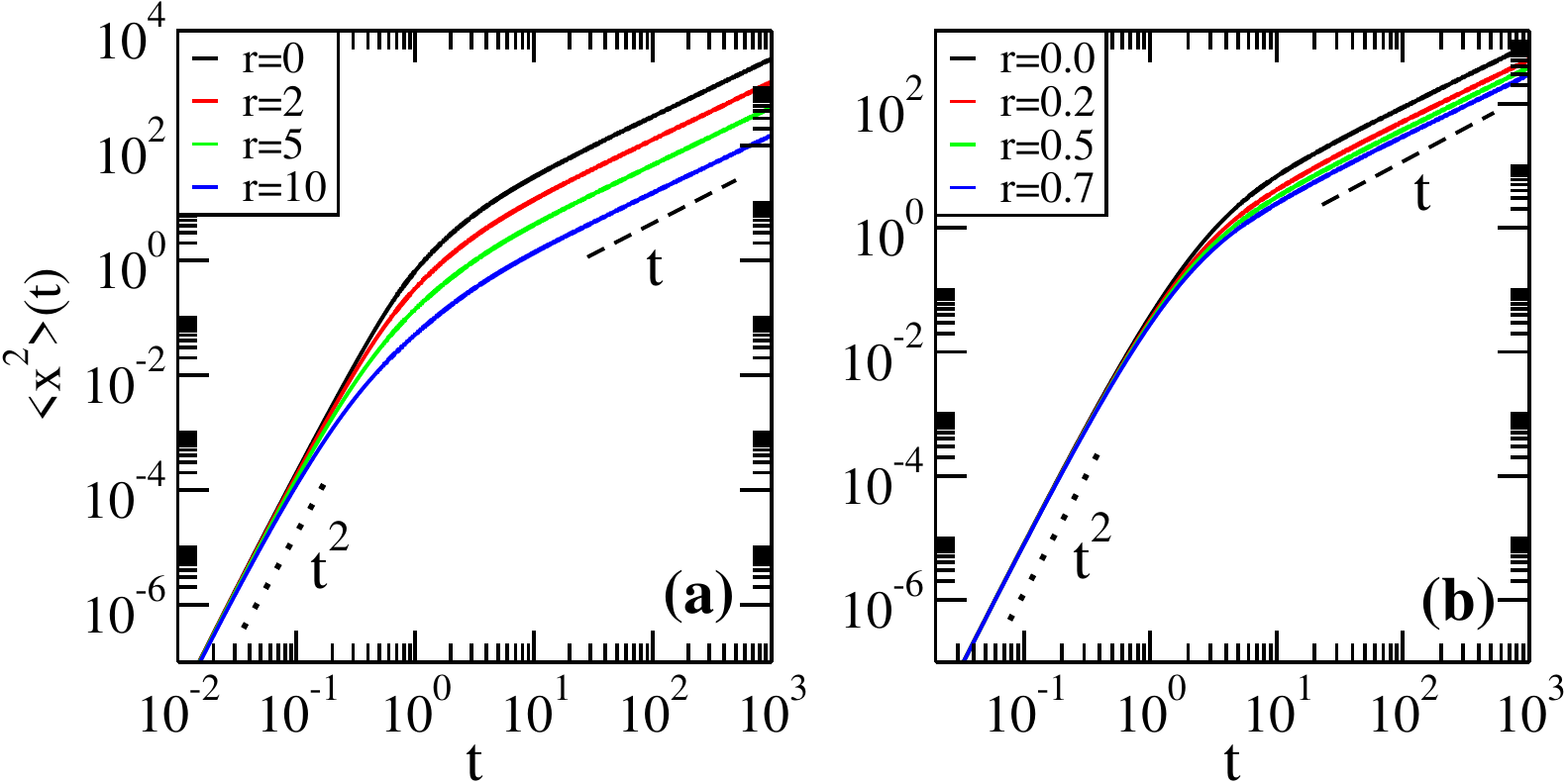}
\caption{\label{p2c4fig5} Plot of mean-squared displacement $\langle x^2\rangle(t)$, as a function of time $t$ for $\lambda=0.5$ at different resetting rates, as indicated. Panel (a) corresponds to the shear-thickening case at $\Sigma=3$, while panel (b) shows the shear-thinning case at $\Sigma=0.6$. The dotted lines labeled $t^2$ represent the ballistic regime, whereas the dashed lines labeled $t$ indicate diffusive behavior.}
\end{figure}

Next, we focus on the transport properties by computing the mean-squared displacement (MSD) $\langle x^2(t)\rangle$. Figure~\ref{p2c4fig5} shows the plot of $\langle x^2(t)\rangle$ versus $t$ on a log-log scale for different values of $r$ at a fixed $\lambda=0.5$: (a) for $g(v)=\tan(v)$ with $\Sigma=3$, and (b) for $g(v)=\tanh(v)$ with $\Sigma=0.6$. In both cases, the particles exhibit ballistic motion at early times with $\langle x^2(t)\rangle \sim t^2$, followed by a crossover to diffusive behavior at longer times,
\begin{eqnarray}
\label{p2c4eqn18}
\langle x^2(t)\rangle=2D_{\rm eff}t.
\end{eqnarray}
Here, $D_{\rm eff}$ is the effective diffusion constant. The resetting rate primarily affects the crossover time but does not alter the asymptotic behavior. The crossover time decreases with increasing resetting rate, independent of the form of $g(v)$ when the other parameters are fixed. Moreover, $\langle x^2(t)\rangle$ decreases with increasing $r$, indicating a reduction in $D_{\rm eff}$. This occurs because the number of particles with velocities $v \approx v_l$ and $v \approx v_u$ decreases as $r$ increases, as evident from Figs.~\ref{p2c4fig2} and \ref{p2c4fig3}. We have verified that the same qualitative behavior persists for other values of $\lambda$ and $\Sigma$.

\begin{figure}
\centering
\includegraphics*[width=0.80\textwidth]{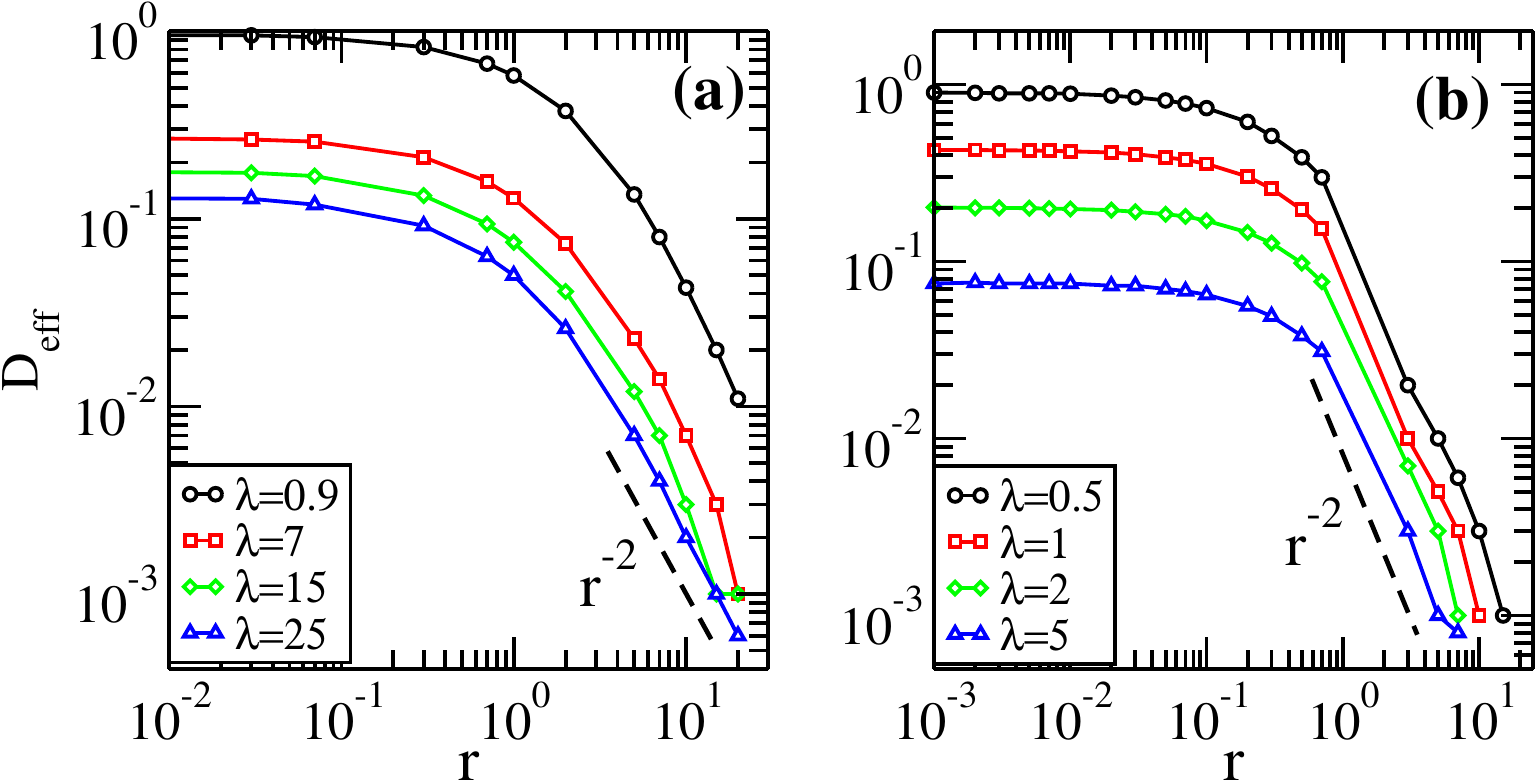}
\caption{\label{p2c4fig6} Log-log plot of the effective diffusivity $D_{\rm eff}$ as a function of the resetting rate $r$ for different flipping rates $\lambda$, as indicated. Panel (a) corresponds to the shear-thickening case with $\Sigma=3$, while panel (b) shows the shear-thinning case with $\Sigma=0.6$. The dotted lines serve as guides to the eye for the $r^{-2}$ behavior.}
\end{figure}

We compute $D_{\rm eff}$ by fitting the numerical data of $\langle x^2(t)\rangle$ versus $t$ using Eq.~(\ref{p2c4eqn18}) in the large-$t$ regime. Figure~\ref{p2c4fig6} presents the variation of $D_{\rm eff}$ with $r$ for different values of $\lambda$ (see the figure caption for details). For fixed activity parameters, $D_{\rm eff}$ shows only a weak dependence on $r$ in the small-$r$ regime. In contrast, for sufficiently large $r$ ($r>1$), $D_{\rm eff}$ decays as $r^{-2}$, irrespective of the specific form of $g(v)$. This can be understood as follows. The mean time interval between two successive resetting events is given by $\langle\Delta t\rangle=r^{-1}$. In the inertia-dominated regime, the average distance traveled by the particle scales linearly with time, i.e., $\langle\Delta x\rangle\sim \langle\Delta t\rangle\sim r^{-1}$. Consequently, the MSD scales as $\langle(\Delta x)^2\rangle\sim r^{-2}$, which in turn implies that $D_{\rm eff}\sim r^{-2}$. This result is the same as that obtained for a Newtonian medium $g(v)\sim v$ with Gaussian noise~\cite{KH24}. A possible explanation is that, at large $r$, particle velocities fluctuate mainly around $v=0$. In this regime, the chosen forms of $g(v)$ also behave as $g(v)\sim v$ for small $v$, which effectively leads to the same power-law behavior irrespective of chosen form of $g(v)$. 

\begin{figure}[ht!]
\centering
\includegraphics*[width=0.70\textwidth]{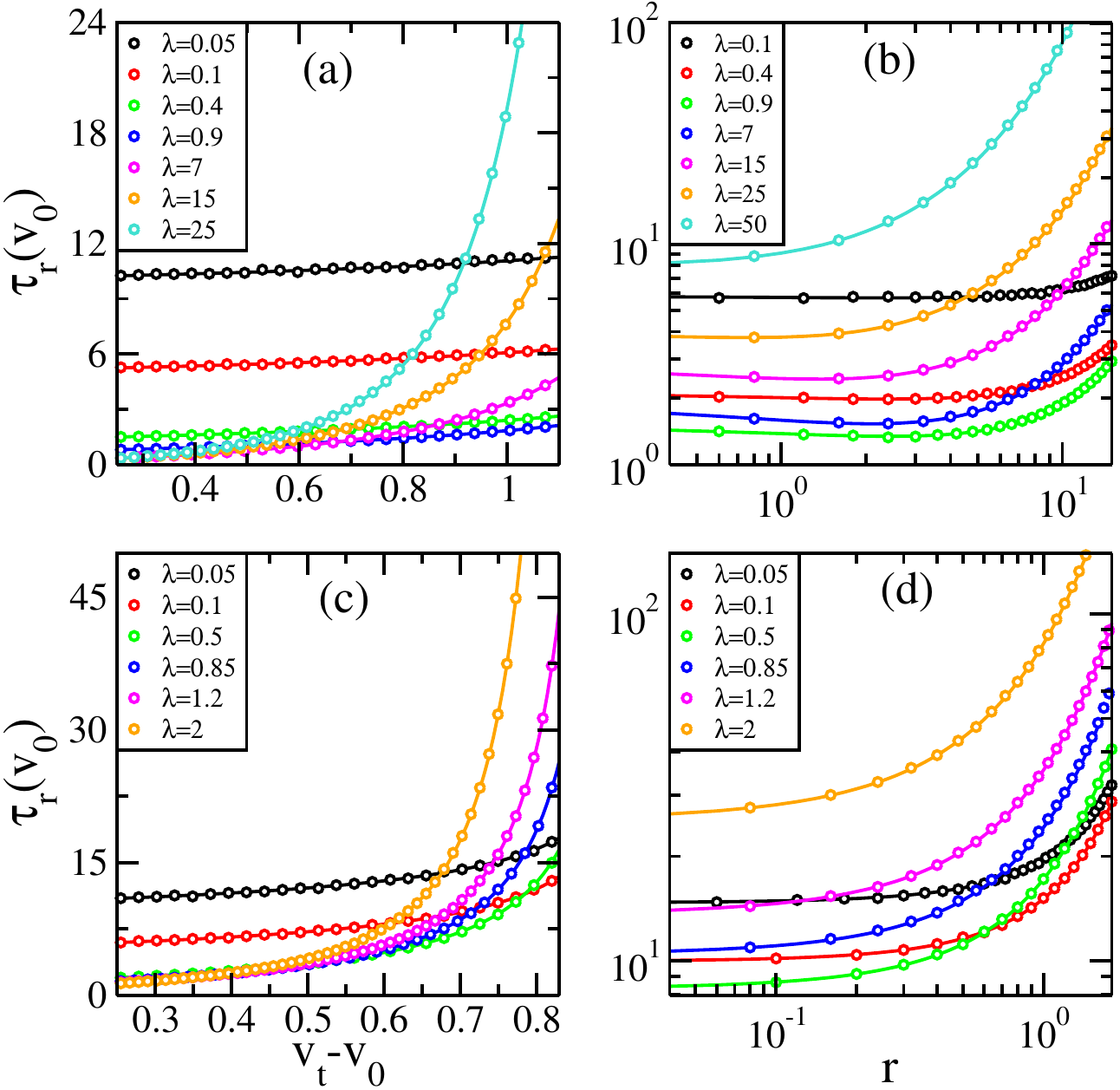}
\caption{\label{fig7} Plot of mean first-passage times, $\tau_r(v_0)$, for non-Newtonian media. (a) Plot of $\tau_r(v_0)$ versus $v_t-v_0$ for $r=2$ and (b) Log-log plot of $\tau_r(v_0)$ versus $r$ for $v_t=0.5$ for the shear-thickening medium with $\Sigma=3$ and different values of $\lambda$, as mentioned. Similarly, (c) Plot of $\tau_r(v_0)$ versus $v_t-v_0$ at a fixed $r=0.2$ and (d) Log-log plot of $\tau_r(v_0)$ versus $r$ at a fixed $v_t=0.5$, for the shear-thinning medium with $\Sigma=0.6$ and different values of $\lambda$, as indicated. At $t=0$, the initial velocity is $v_0=-0.249$, and the velocity of resetting is $v_r=0$. In all panels, data from particle-based simulations are shown as points, while the numerical results obtained by solving Eqs.~\eqref{neqn26} and \eqref{neqn27} are represented by solid lines of the same color.}
\end{figure}

Next, we present the results of MFPT $\tau_r(v_0)$ corresponding to our model. We obtain $\tau_r^\pm(v_0)$ by solving the coupled differential equations \eqref{neqn26} and \eqref{neqn27} numerically using a \textit{finite-difference upwind scheme}~\cite{p2c1pmazumdarcomp21}. We discretize the velocity space using a uniform mesh size $\Delta v_0=1.5\times10^{-4}$. Starting from arbitrary initial values
for $\tau_r^\pm(v_0)$s close to zero, Eqs.~\eqref{neqn26} and \eqref{neqn27} are iterated until the maximum difference between successive iterations is less than $10^{-10}$. In the simulation, we set the initial velocity to $v_0=-0.249$ with the reset velocity $v_r=0$ and vary the target velocity $v_t$ and the rate of resetting $r$. Figure~\ref{fig7}(a) shows the variation of $\tau_r(v_0)$ as a function of the velocity difference between the target velocity $v_t$ and the initial velocity $v_0$, i.e., $(v_t - v_0)$, for a shear-thickening medium at fixed $r=2$, $\Sigma=3$, and different values of $\lambda$. We observe that, for larger values of $\lambda$, $\tau_r(v_0)$ initially takes relatively small values but then increases rapidly and eventually diverges as $(v_t-v_0)$ increases. This behavior arises because frequent velocity flipping reduces the probability of reaching $v_t$, located far from $v_0$. In contrast, for smaller values of $\lambda$, $\tau_r(v_0)$ starts from relatively larger values and remains nearly constant over the entire range of $(v_t-v_0)$. In this regime, the velocity dynamics is more persistent, and the fixed resetting rate does not significantly affect the approach to the target. Typically, $\tau_r^+(v_0)$ is small, whereas $\tau_r^-(v_0)$ is large and therefore makes a dominant contribution to $\tau_r(v_0)$, since a particle subject to an active force $-\Sigma$ cannot reach $v_t$ in the positive velocity domain until the active force flips sign. Figure~\ref{fig7}(b) shows the plot of $\tau_r(v_0)$ as a function of $r$, at $v_t=0.5$ and various $\lambda$ in the same media. For small to moderate $\lambda$ (typically $\lambda<1$) in the small-$r$ regime, $\tau_r(v_0)$ decreases gradually with $r$. However, upon further increasing $r$, $\tau_r(v_0)$ rises rapidly, indicating the existence of an optimal resetting rate. At a large value of flipping rate($\lambda$), for example $\lambda=50$, we did not find any optimal resetting rate as $\tau_r(v_0)$ increases monotonically with $r$.

Figure~\ref{fig7}(c) is analogous to  Fig.~\ref{fig7}(a), but for the shear-thinning medium at a fixed resetting rate $r=0.2$. Apart from minor quantitative differences, the two plots exhibit qualitatively similar behavior, indicating that the underlying mechanism remains essentially the same and does not depend strongly on the nature of the medium, i.e., the specific form of $g(v)$. In Fig~\ref{fig7}(d), we plot $\tau_r(v_0)$ as a function of $r$ on a log-log scale for the shear -thinning case with $v_t=0.5$, analogous to Fig.~\ref{fig7}(b). Unlike the shear-thickening case, where an optimal resetting rate exists for small and intermediate values of $\lambda$, no such optimal rate is observed here. Instead, $\tau_r(v_0)$ increases monotonically with $r$ for all values of $\lambda$.

\section{Conclusion}\label{sec4}
In summary, we have studied the dynamics of an athermal IRTP moving through a non-Newtonian medium under a velocity-resetting protocol in $d=1$. The viscous drag force acting on the particle is modeled by a nonlinear function $g(v)$, whose functional form depends on the nature of the medium. In the case of a shear-thinning medium, $g(v)\sim \tanh(v)$, whereas for a shear-thickening medium, $g(v)\sim \tan(v)$. The intrinsic activity of the IRTP is modeled by a symmetric dichotomous noise with strength $\Sigma$ and flipping rate $\lambda$.
We then write down the FP equation for the velocity distribution function $P(v,t)$, which, in the presence of a constant resetting rate $r$, does not admit a closed-form solution even in the steady state. Therefore, we compute the steady-state velocity distribution $P_s(v)$ numerically by solving the FP equation and also from particle-based simulations. The results obtained from both methods are in excellent agreement. For fixed values of $\lambda$ and $\Sigma$, when $r$ is varied, $P_s(v)$ undergoes several intriguing transitions for both the shear-thinning and shear-thickening cases. In particular, we observe a cusp-like singularity at $v=0$ for sufficiently large $r$; further increasing $r$ sharpens the cusp in both cases. We calculate the mean-squared velocity for different values of $r$ for the above choices of $g(v)$ using both methods, and the results are again in excellent agreement at long times. We find that the particles exhibit diffusive motion at long times. From the slope of the mean-squared displacement versus $t$ curve, we obtain the effective diffusion constant $D_{\rm eff}$, which decays as $r^{-2}$ in the large-$r$ limit. This behavior is the same as that observed for a Newtonian medium with Gaussian noise, suggesting that for large resetting rates $D_{\rm eff}$ does not depend on the nature of the underlying medium. 

Further, we investigate the effect of stochastic resetting on the mean first passage time $\tau_r(v_0)$ of a particle in non-Newtonian media. Using the survival equation formalism, we derive a second-order inhomogeneous differential equation for $\tau_r(v_0)$. However, obtaining an analytical closed-form expression for $\tau_r(v_0)$ from this differential equation is not feasible for the chosen forms of $g(v)$, and also due to the absence of appropriate boundary conditions. Therefore, we compute $\tau_r(v_0)$ numerically by solving a pair of coupled differential equations for which boundary conditions are available, and compare the results with particle-based simulations. We find that $\tau_r(v_0)$ depends strongly on the nature of the medium. In shear-thickening media, $\tau_r(v_0)$ exhibits a non-monotonic dependence on the resetting rate $r$. For small and moderate values of $\lambda$, $\tau_r(v_0)$ initially decreases in the small-$r$ regime. As $r$ is increased further, $\tau_r(v_0)$ grows rapidly, indicating the existence of an optimal resetting rate that minimizes the time required for the particle to reach a prescribed target velocity $v_t$. However, for large $\lambda$, this non-monotonic behavior disappears, and $\tau_r(v_0)$ increases monotonically with $r$. This is in sharp contrast to the shear-thinning case, where $\tau_r(v_0)$ increases monotonically with the resetting rate $r$ for all values of $\lambda$.

\ \\
\noindent{\bf Acknowledgments:} S. Howlader and S. Mondal are grateful to IISER-Mohali, India for the financial support through a senior research fellowship.

\ \\
\noindent{\bf Conflict of interest:} The authors have no conflicts to disclose.

\end{document}